\documentclass[
    ,final            
  ]
  {aipproc}

\layoutstyle{6x9}

\def \g {$\gamma$}

\usepackage{natbib}

\begin{document}

\title{Constraints on Extragalactic Background Light from Cherenkov telescopes: status and perspectives for the next 5 years}


\author{Daniel Mazin}{
  address={Institut de F\'{i}sica d'Altes Energies, Edifici Cn. UAB, E-08193 Bellaterra (Barcelona), Spain}
}



\begin{abstract}
Very high energy (VHE, E > 30 GeV) \g-rays are absorbed via interaction with low-energy photons from the 
extragalactic background light (EBL) if the involved photon energies are above the threshold for electron-positron pair creation. The VHE \g-ray absorption, 
which is energy dependent and increases strongly with redshift, distorts the VHE energy spectra observed from distant objects. 
The observed energy spectra of the AGNs carry therefore an imprint of the EBL. Recent detections of hard spectra of distant 
blazars (z = 0.11 -- 0.54) by H.E.S.S. and MAGIC put strong constraints on the EBL density in the optical to near infrared waveband. It is, however, not yet
possible to distinguish between an intrinsic softening of blazar spectra and a softening caused by the interaction with low energy
EBL photons. In this paper, we give an overview of the EBL constraints, their limitations and perspectives for the joint
efforts of the Fermi Gamma-Ray Space telescope and imaging atmospheric Cherenkov telescopes.
\end{abstract}

\maketitle


\section{Introduction}

During the star and galaxy formation history a diffuse extragalactic radiation
field has been accumulated in the ultraviolet to far infrared wavelength regimes.
This radiation field, commonly referred to as the extragalactic background
light (EBL), is the second largest, in terms of the contained energy,
background after the Cosmic Microwave Background of 2.7 K (CMB). 
While the CMB conserves the structure of the universe at the moment of the decoupling of
matter and radiation following the Big Bang (at redshift z $\approx$ 1000), the EBL is a calorimetric measure of the entire 
radiant energy released by processes of structure formation that have occurred since the decoupling. 

\begin{figure}[htb]
\begin{minipage}{0.48\textwidth}
\centering
\includegraphics[width=\textwidth]{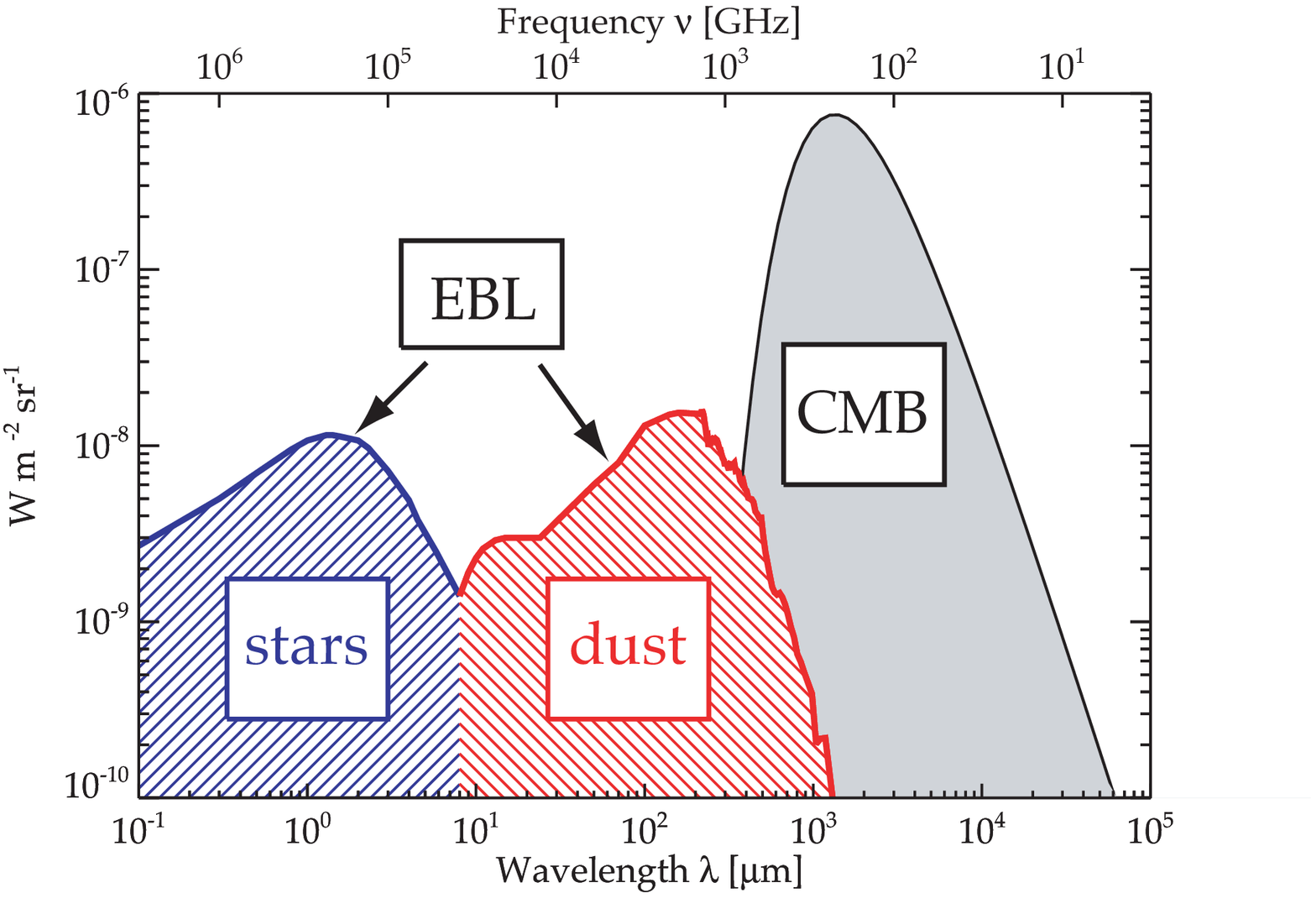}
\end{minipage}
\hfill
\begin{minipage}{0.48\textwidth}
\centering
\includegraphics[width=\textwidth]{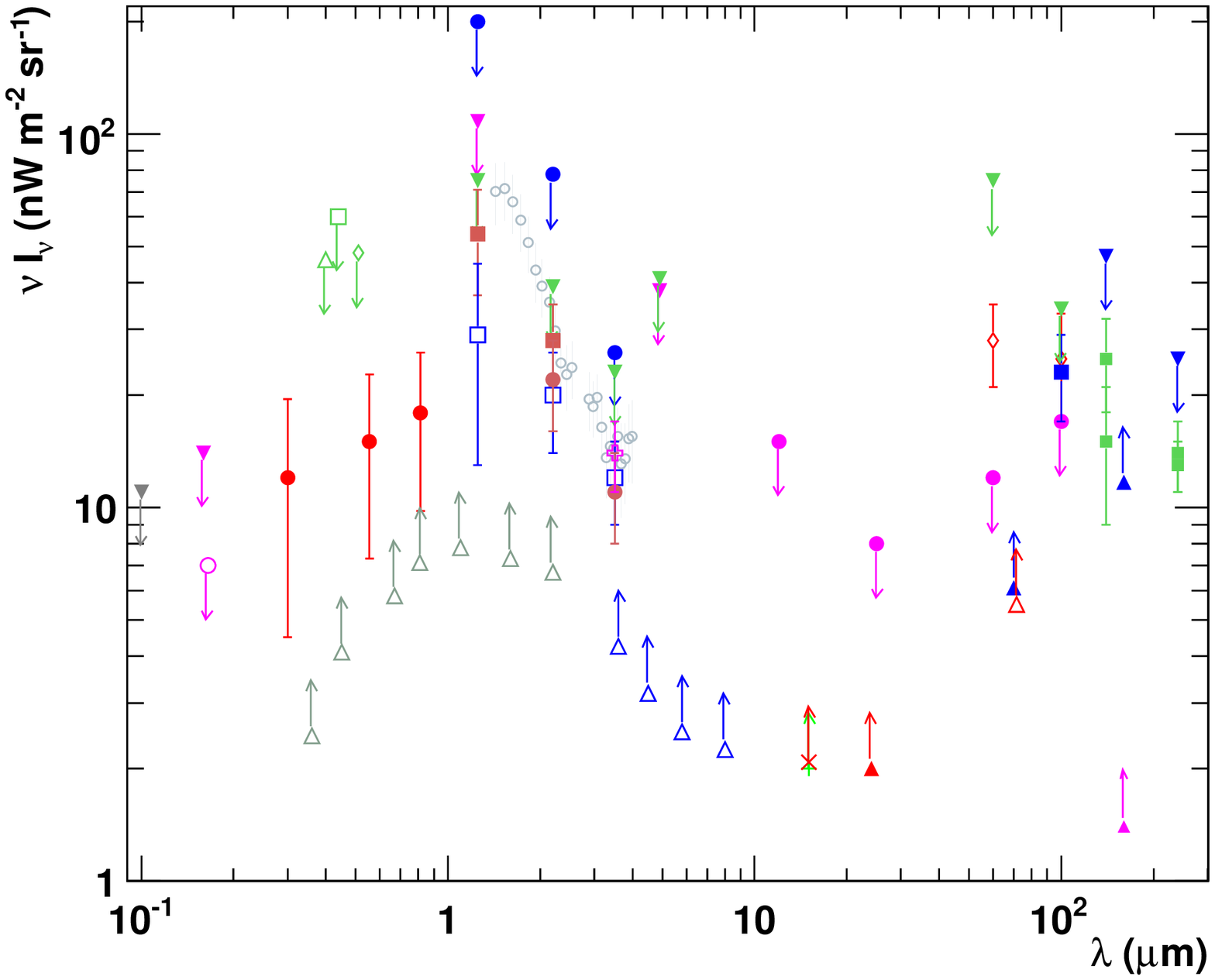}
\label{fig:ebl_sketch} 
\caption
{
   \textit {Left: }Schematic Spectral Energy Distributions (SED) of the most
   important (by intensity) backgrounds in the universe.
   From right to left: the cosmic microwave background (CMB),
   the cosmic infrared background (CIB) and the cosmic optical
   background (COB). The last two components together are called EBL. 
   Plot adopted from \cite{dole:2006a}.  
   \textit {Right: } 
{EBL measurements and limits (status end 2006).
Tentative detection in the UV/optical:
\cite{bernstein:2002a,bernstein:2005a} (filled red circle);
Lower limits from galaxy counts: 
\cite{madau:2000a} (open grey triangles),
\cite{fazio:2004a} (open blue triangles);
Detections in the near IR:
\cite{dwek:1998b} (open pink cross),
\cite{gorjian:2000a} (filled brown circle),
\cite{wright:2000a} (open blue squares),
\cite{matsumoto:2005a} (small open grey circles).
Other symbols see in \cite{mazin:2007a}.}
}
\end{minipage}
\end{figure}

A closer look to the UV -- infrared backgrounds is given in
Figure~\ref{fig:ebl_sketch}, left plot.  From right to left, the spectral energy
distributions of the three major components are shown: the cosmic microwave
background (CMB), the cosmic infrared background (CIB), and the cosmic optical
background (COB).  
The COB, peaking at around 1$\,\mu$m is believed to originate directly from stars.
The CIB, having its peak at $\sim$100$\,\mu$m, results mostly from
starlight that has been absorbed by dust inside galaxies and reemitted at higher wavelengths.  
Throughout this paper, we will refer to COB and CIB together as EBL.
Other contributions, like emission from AGN
and quasars are expected to produce no more than 5 to 20\% of the total EBL
density in the mid IR (see e.g. \cite{matute:2006a} and references therein).

The EBL is difficult to measure directly due to strong foregrounds from our solar system
and the Galaxy. 
The observation of distant sources of VHE $\gamma$-rays using
Imaging Air Cherenkov Telescopes (IACT, such CANGAROO, H.E.S.S., MAGIC or VERITAS) provides a unique indirect
measurement of the EBL (see below).
The precision of the EBL constraints set by the IACT improved remarkably in the last few years.
Contemporaneously with the IACT constraints, 
there has been rapid progress in resolving a significant fraction of this background with
the deep galaxy counts at infrared wavelengths from the Infrared Space
Observatory ({\it ISO}) and from the  \textit{Spitzer} satellite 
as well as at sub-millimeter wavelengths from the
Submillimeter Common User Bolometer Array (SCUBA) instrument. 
The current status of direct and indirect EBL measurements
(excluding limits from the IACTs) is shown in Fig.~\ref{fig:ebl_sketch}, right plot. 

\begin{figure}[htb]
\includegraphics[width=0.45\textwidth]{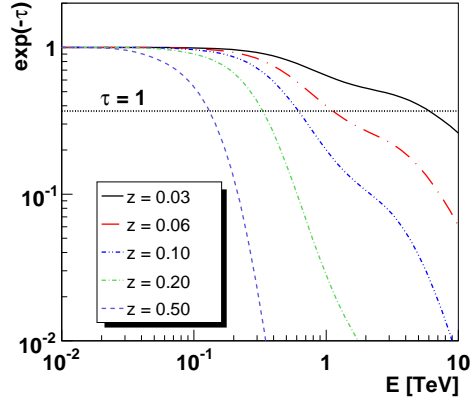}
\caption
{\label{fig:atten} Attenuation coefficients $\exp(-\tau)$ as a function of \g-ray energy.
The ``best fit'' model from \cite{kneiske:2002a} is used for the EBL density. The curves represent 
the expected attenuation of the VHE \g-ray spectra for different source distances.
The horizontal dotted line corresponds to the optical depth  $\tau = 1$, i.e. it crosses the attenuation
curves at the energy, above which the universe becomes opaque.}
\end{figure}

In total, the collective limits on the EBL
between the UV and far-IR confirm the expected two peak structure, although the
absolute level of the EBL density remains uncertain by a factor of 2 to 10. 
In addition to this consistent picture, several direct 
measurements in the near IR have also been reported  \citep[e.g.][]{matsumoto:2005a}, 
significantly exceeding the expectations from source counts 
(see \cite{hauser:2001a} and \cite{Kashlinsky2005:EBLReview}
for recent reviews).  If this claimed excess of the EBL is
real, it might be attributed to emissions by the first stars in the history of
the universe. These so-called Population III stars are believed to be heavy stars
with a very low metallicity. 

This paper is structured as follows. In the next section we will describe the absorption
effect on spectra of distant VHE $\gamma$-ray sources as blazars due to the EBL 
and a possibility of using measured blazar spectra to constrain the EBL.
Then we present the current status of the EBL constraints using blazar spectra and discuss
their validity. Finally, we sketch possible improvements of the EBL measurements in the next five years using the new generation of the IACTs and the Fermi satellite.

\section{VHE \g-rays from blazars as a probe of the EBL}

On the way from the source to the observer,
VHE \g-rays can suffer absorption losses by interaction with the low energy photons of the EBL.
The corresponding reaction is the pair-production of an electron-positron pair:
\begin{equation}
  \gamma_{\mbox{\tiny{VHE}}} + \gamma_{\mbox{\tiny{EBL}}} \longrightarrow e^+ + e^- \;\;\;\; 
      \mathrm{with} \;\;E_{\gamma_{\,\mathrm{VHE}}} \cdot E_{\gamma_{\,\mathrm{EBL}}} > (m_e c^2)^2
\end{equation}

The optical depth of the VHE \g-rays, $\tau(E)$, emitted at the redshift $z$,
can then be calculated by solving the three-fold integral (see also \cite{dwek:2005a}):
  \begin{eqnarray}
    \label{eq:tau1}
      \tau(E_{\gamma},\,z) & = & \int_{0}^{z} {\rm d}\ell ({\rm z'}) \, \int_{-1}^{1} d\mu\,\frac{1-\mu}{2} 
                        \int_{\epsilon'_{th}}^{\infty} d\epsilon'\,n(\epsilon',z') \,\sigma_{\gamma\gamma}(\epsilon',E',\mu) \\
       \mu & : \, = & \cos{\theta} \nonumber \\
    n\left(\epsilon\,\right) & :\, = & \, \mbox{EBL energy density} \nonumber \\
    {\rm d}\ell ({\rm z})   & :\, = & \, \mbox{distance element} \nonumber
  \end{eqnarray}
The expected optical depth for VHE \g-ray fluxes from sources 
at different redshifts is shown in
Fig.~\ref{fig:atten}.  Hereby, the ``best fit'' model from \cite{kneiske:2002a}
is used to estimate the EBL density.  One can see that the optical depth is not
only redshift but also energy dependent. 
The crossing points of the
horizontal dotted line in Fig.~\ref{fig:atten} ($\tau = 1$) with the
attenuation lines define then the energy for the different  redshifts, at which
the universe becomes opaque. 
It is evident that a low threshold of a VHE \g-ray
detector is essential to observe distant sources. 

There are two major aspects concerning the interconnection between VHE \g-rays and the EBL: 
\begin{enumerate}
  \item The EBL leaves a unique imprint on the VHE spectra. This imprint can be used to study the EBL.
  \item The impact of EBL onto VHE spectra is fundamental. 
        We need to understand EBL in order to study intrinsic properties
        of the VHE \g-ray sources.
\end{enumerate}
Concerning point (1): From a single observed energy spectrum of a distant VHE \g-ray source, 
it is rather difficult if not impossible to distinguish between the imprint of the EBL
and intrinsic features of the source. Observed features can be source inherent due 
to an internal absorption inside the source or due to a source, which does not provide necessary conditions 
for acceleration of charged particles to high enough energy.
A way to pin-point the EBL impact is to use population studies of many extragalactic sources:
whereas the intrinsic features might be different, the imprint of the EBL is the same (at a given redshift).
With the current population of VHE \g-ray sources, it is only possible to set limits on the EBL, arguing
that the observed spectra contain at least the imprint of the EBL.

Note that only distant extragalactic VHE \g-ray emitters suffer from the absorption by the
EBL. For the galactic sources, the effect is negligible up to energies of about
100 TeV. For higher energies, the absorption by the photon field of the 
CMB starts to be important.

\section{Status of the EBL limits set by Cherenkov telescopes}

The observed VHE \g-ray spectrum of a source ($F_{\mathrm{obs}}$) can be used to 
reconstruct the intrinsic spectrum ($F_{\mathrm{intr}}$), i.e.\ the one at the source location:

\begin{equation}
  F_{\mathrm{intr}} = \exp(\tau) \times F_{\mathrm{obs}},
\end{equation}
whereas $\tau$ is an energy and distance dependent optical depth of VHE \g-rays.
By measuring the observed spectrum of a source and inferring certain limits on the intrinsic spectrum of the source, 
it is thus possible to constrain $\tau$ and, therefore, to constrain the EBL density.

Flat Spectrum Radio Quasars (FSRQ) and BL Lacs are subclasses of AGNs, which have
their jet aligned close to the line of sight of the observer. The two
subclasses are jointly called ``blazars''.
According to an accepted paradigm, in blazars VHE \g-ray photons are produced
by VHE electrons or protons, which are accelerated to VHE energies through
a shock acceleration. High bulk motion Lorentz factors in the jets and the jet alignment
towards the observer enhance chances to detect VHE \g-ray emission.
Blazars are indeed the so far most prominent class of VHE \g-ray emitting extragalactic objects.
In the shock acceleration models, 
the hardest index obtained for the accelerated particles
is $s = 1.5$ (see e.g. \cite{malkov:2001a}). 
In the case of protons interacting with ambient plasma, 
the resulting $\gamma$-ray spectrum has the same slope as $s$, i.e.\ $\Gamma_{\rm int} = 1.5$. 
In the case of electrons, 
the spectrum of the $\gamma$-rays emitted through inverse Compton
scattering is expected to be steeper than $1.5$ 
under most circumstances. Therefore, different authors \cite{aharonian:2003a,aharonian:2006:hess:ebl:nature,albert:2008:3C279:science} assumed  $\Gamma_{\rm int} = 1.5$ 
to be the hardest possible intrinsic spectrum, and using this criterion 
stringent EBL limits were derived. It was, however, argued  
that stochastic electron acceleration \cite{stecker:2007a}, 
truncated electron spectra \cite{katarzynski:2006a} 
or internal absorption (e.g. \cite{sitarek:2007a,aharonian:2008a})
can lead to even harder VHE \g-ray spectra than with an index of $\Gamma_{\rm int} = 1.5$
although up to now no harder spectra with $\Gamma_{\rm int} < 1.5$ have been observed at lower energies, where no EBL absorption can take place.

\subsection{H.E.S.S. limits on the EBL}

\begin{figure}
\includegraphics[viewport=20 50 585 700,clip,width=0.4\textwidth]{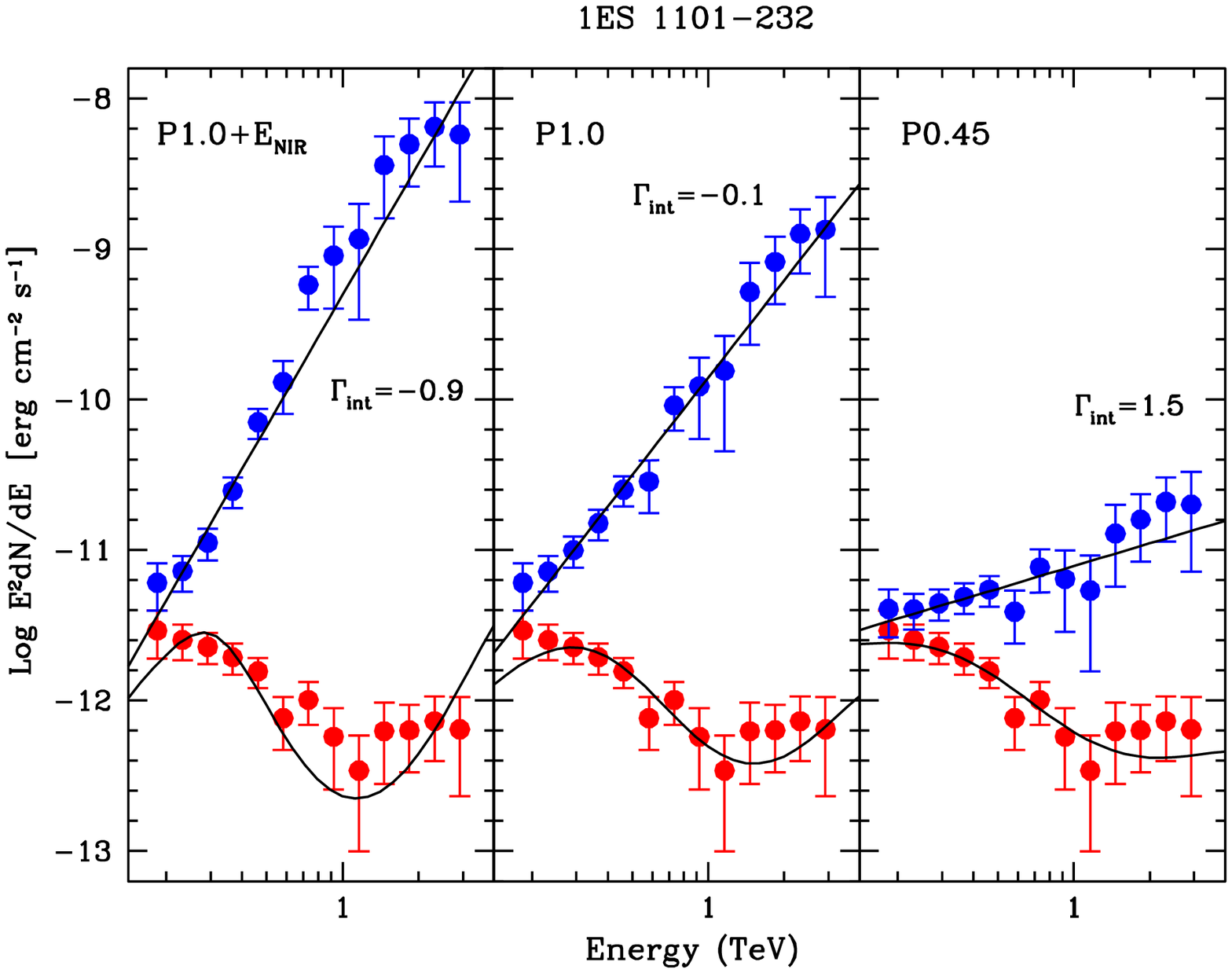}
\includegraphics[width=0.4\textwidth]{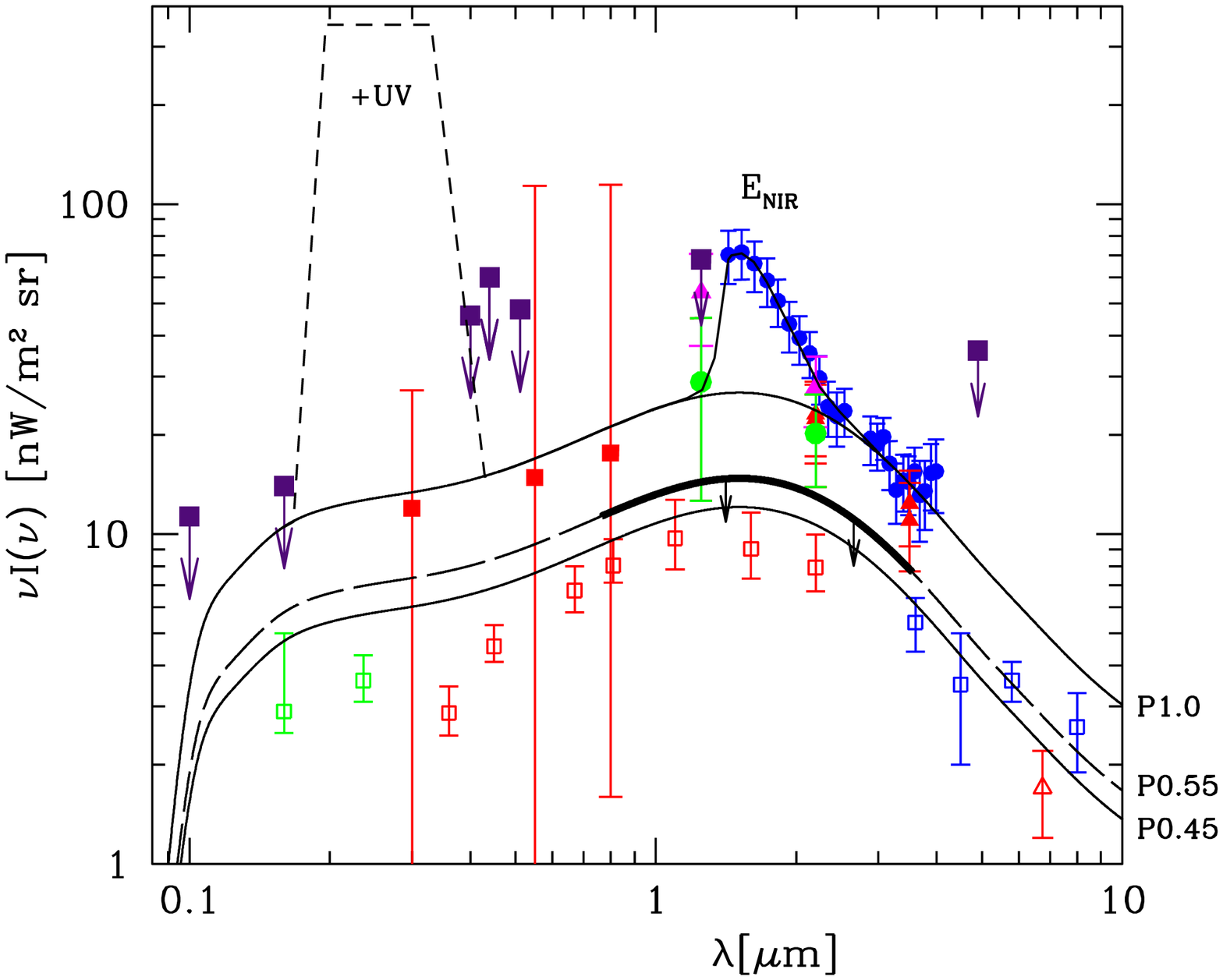}
\caption
{\label{fig:hess} HESS 
{\it Left:} The H.E.S.S.\ time-averaged spectrum of 1ES\,1101-232 (red)
together with a reconstructed intrinsic spectrum (blue) for three different
EBL shapes. The first two EBL shapes lead to an unnaturally hard
intrinsic spectrum (left and middle).
{\it Right:}
SED of the EBL. 
The thick black line between 0.8 and 4\,$\mu$m shows 
the H.E.S.S.\ limit.
In the long-dashed regions, higher fluxes than the limit model
would not be in conflict, as long as the 
flux in the 1-3\,$\mu$m range is within or around the limit. Figures from
\cite{aharonian:2006:hess:ebl:nature}.} 
\end{figure}

The H.E.S.S. collaboration published a detection of two intermediate redshift
blazars 1ES\,1101-232 (z = 0.186) and H\,2356-309 (z = 0.165) 
\citep{aharonian:2006:hess:ebl:nature}. 
Both observed spectra (measured in the range 150 GeV -- 3 TeV) show a relatively hard
spectral index of 2.9 and 3.1, respectively. Using the criterion that the intrinsic
blazar spectrum cannot be harder than $\Gamma_{\rm int} = 1.5$, the authors derived a 
stringent upper limit on the EBL density in the region between 0.8 and 4 $\mu$m 
(see Fig.~\ref{fig:hess}).  
The derived upper limits imply a low level EBL density in agreement with the 
expectations from standard galaxy evolution models.
The limits, in turn, rule out a cosmological origin of the near infrared excess 
\citep[e.g.][]{matsumoto:2005a}. 

Using these EBL limits, physical parameters of Population III stars were explored
by \cite{raue:2008a}. Sensitive limits on their star formation rate, 
metallicity and initial mass function were derived, which are comparable with 
or better than limits derived by other means 
(e.g.\ methods to derive the epoch of cosmic reionization).

\subsection{Combined EBL limits using all blazars until 2006}

\begin{figure}
\includegraphics[width=0.4\textwidth]{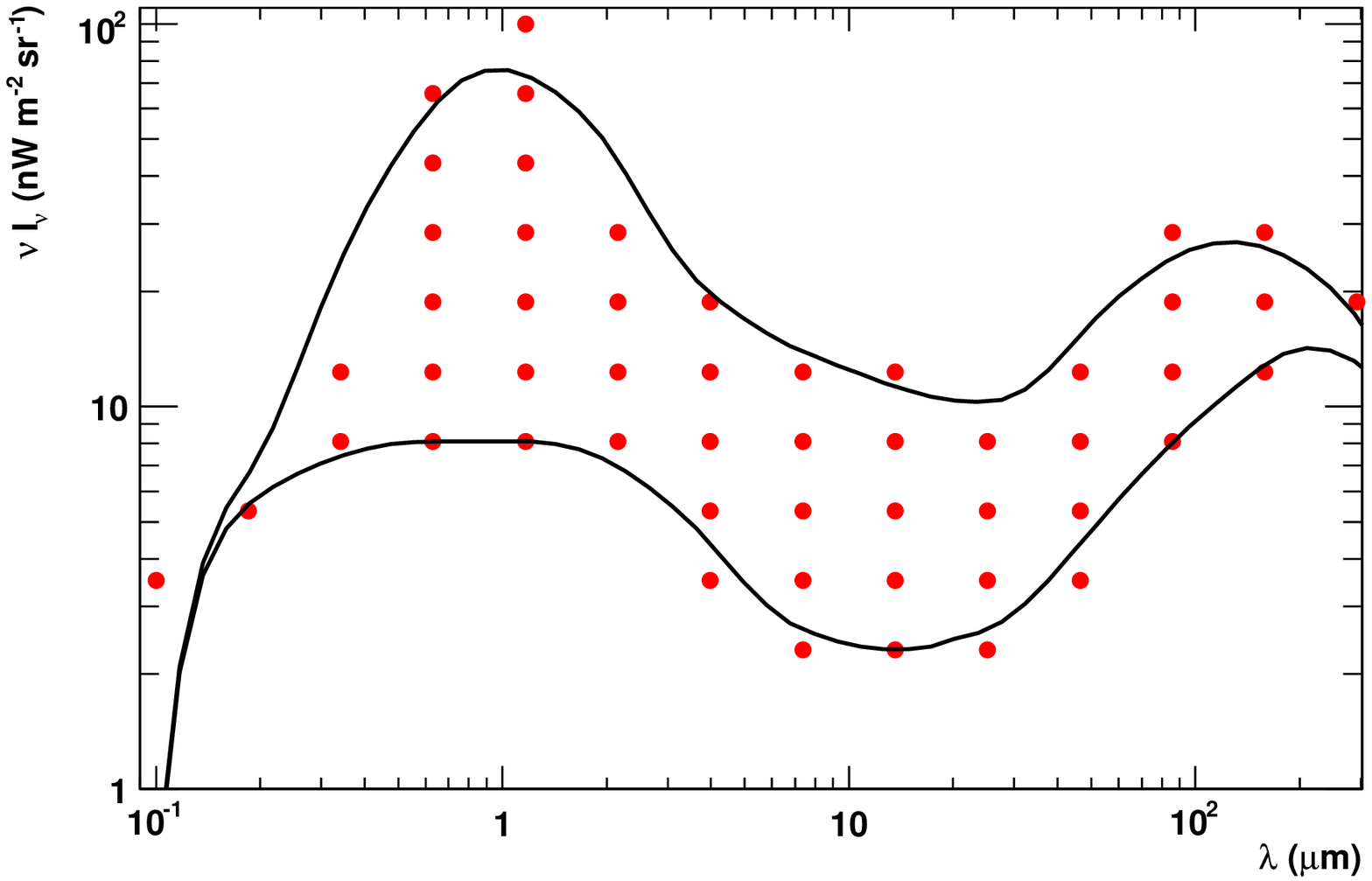}
\includegraphics[width=0.4\textwidth]{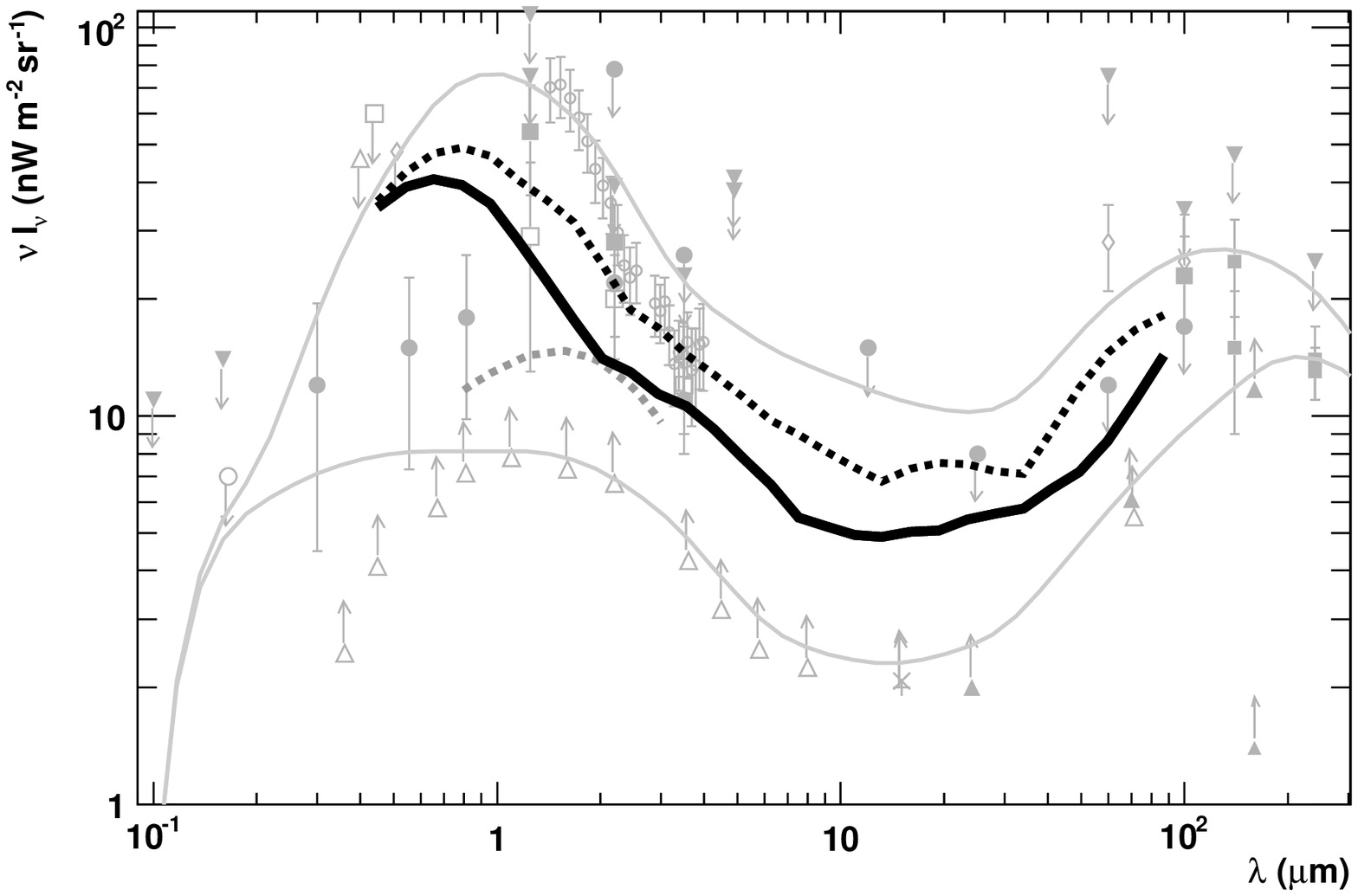}
\caption
{\label{fig:combined} Combined results.
\textsl{Left:} Grid points (red filled circles) and minimum and maximum shape of the scan.
In total,  8$\,$064$\,$000 hypothetical EBL shapes were tested.
\textsl{Right:} 
Combined results from the \textit{extreme} scan 
(dashed black line) in comparison to the result from the 
\textit{realistic} scan (solid black line). 
Grey lines are the minimum and the maximum shapes tested in the scan. 
Grey markers are direct measurements and limits.
Figures from \cite{mazin:2007a}.
} 
\end{figure}

A common criticism of the EBL limits derived as shown above is that they use 
only few blazars (therefore not providing consistency with other sources) 
and that the limits are obtained by 
assuming a certain EBL model and e.g.\ scaling it, or by exploring just a few
details, i.e.\ the derived limits become very model-dependent. In order to
avoid this dependency \cite{mazin:2007a} performed a scan over many hypothetical 
EBL realizations (over 8 million different ones). 
The authors also tested all available blazar spectra
(until 2006) to generalize the EBL limits.
The scanned EBL region is shown in Fig.~\ref{fig:combined}, left plot. The derived upper 
limits on the EBL density are shown in the right plot of  Fig.~\ref{fig:combined}.
Two limits are shown: the solid line represents the upper limit assuming that
the intrinsic blazar spectrum cannot be harder than $\Gamma_{\rm int} = 1.5$, whereas
the dashed line shows the limit for  $\Gamma_{\rm int} = 2/3$. The latter one can be
understood as the most conservative one as it is derived for monoenergetic 
electrons, which are responsible for the inverse Compton scattering of ambient photons.
One can see that the derived limits favor a low EBL level and are in good agreement 
with galaxy counts from the optical to the mid infrared regimes. Again, the 
cosmological origin of the near infrared excess \citep[e.g.][]{matsumoto:2005a}
can be ruled out even for the extreme case of  $\Gamma_{\rm int} = 2/3$.

\subsection{MAGIC limits on the EBL}

\begin{figure}[ht]
\includegraphics[width=0.4\textwidth]{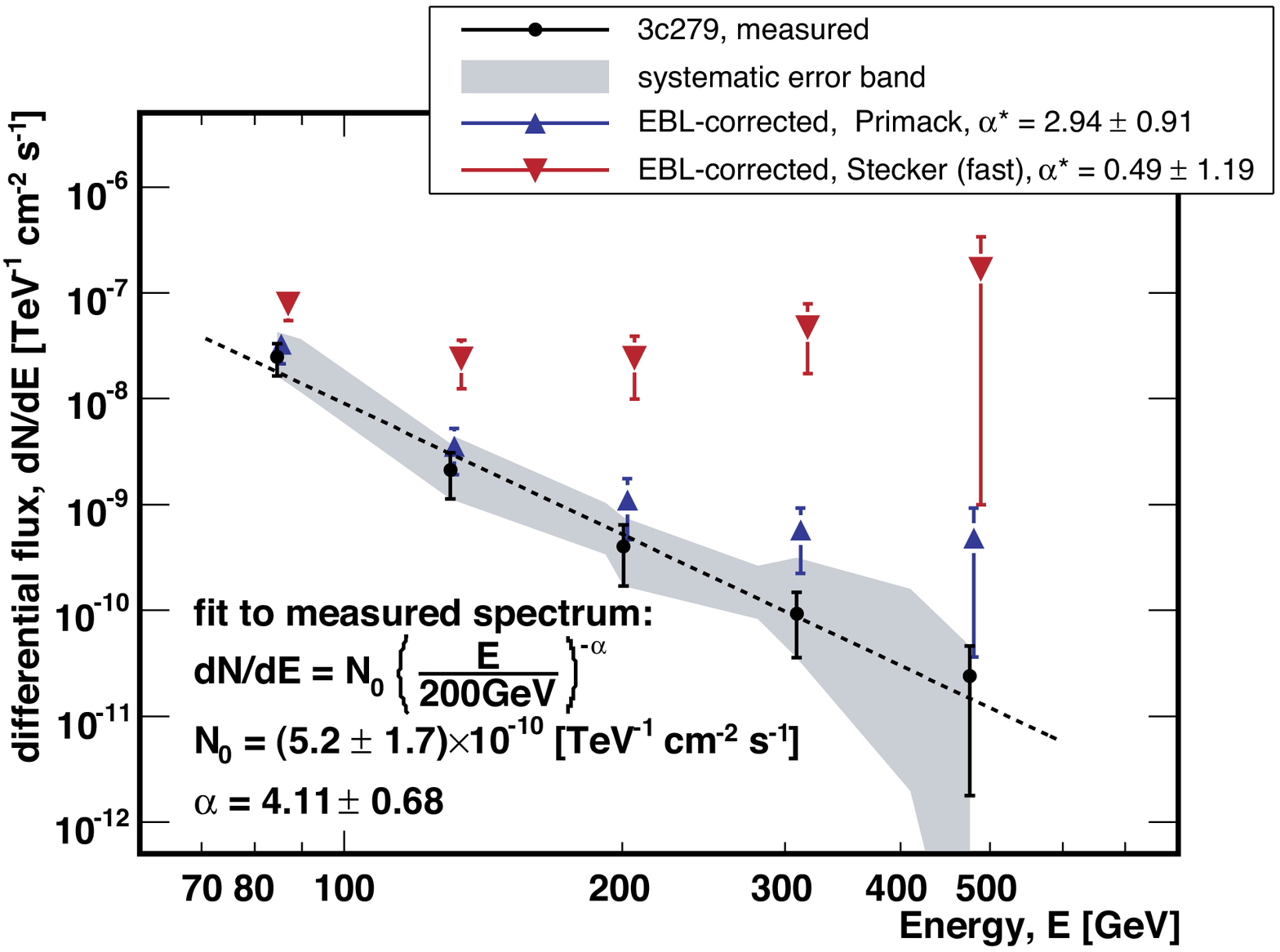}
\includegraphics[width=0.4\textwidth]{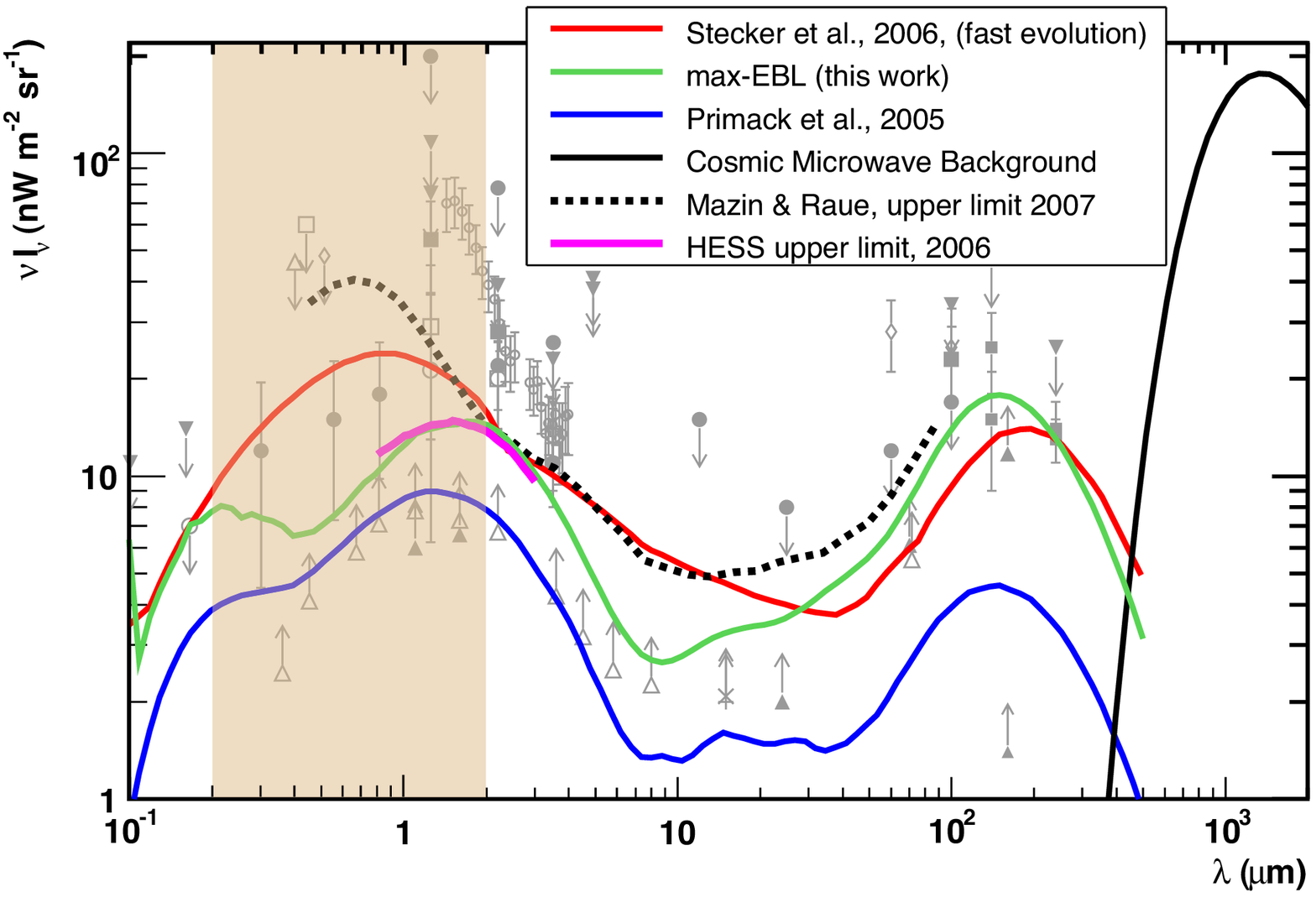}
\caption
{\label{fig:magic} MAGIC
\textsl{Left:} 
Spectrum of 3C~279 measured by MAGIC.
The grey area includes the combined statistical ($1\sigma$) and
systematic errors. 
The dotted line shows the result of a power law fit.
The blue and red triangles are measurements
corrected on the basis of two models for the EBL density: \cite{primack:2005a} (blue)
and \cite{stecker:2006a} (red).
\textsl{Right:} 
SED of the EBL. Some of the EBL models are shown: 
 \cite{primack:2005a} (blue),  \cite{stecker:2006a} (red) and \cite{albert:2008:3C279:science}
(green). The latter one corresponds to the derived EBL limit.
The shaded vertical band indicates the range of
frequencies where the MAGIC measurement is most sensitive.
}
\end{figure}

in 2007, the MAGIC collaboration reported a detection of a very distant (z = 0.536)
radio quasar 3C\,279 at energies above 80 GeV \cite{teshima:2007b,albert:2008:3C279:science}. 
The measured energy spectrum of 3C\,279 
extends up to $\approx$500\,GeV (see Fig.~\ref{fig:magic}, left plot), 
which implies a very low EBL level. 
In order to derive an EBL limit, the MAGIC 
collaboration used a realistic EBL model of \cite{kneiske:2002a}. The authors 
\cite{albert:2008:3C279:science} fine-tuned 
physical parameters of the EBL model in order to comply  
with the requirement that the intrinsic spectrum of 3C\,279 cannot be harder 
than  $\Gamma_{\rm int} = 1.5$. The resulting maximum allowed EBL model is shown 
by the green line in  Fig.~\ref{fig:magic}, right plot. The EBL limit derived in
\cite{albert:2008:3C279:science}
not only confirms limits from \citep{aharonian:2006:hess:ebl:nature} and 
\cite{mazin:2007a} but also probes for the first time the EBL 
at higher redshifts $0.2 < $ z $ < 0.5$. 
Moreover, due to a low energy threshold of MAGIC, the limit extends 
into the ultraviolet regime: the EBL region between 0.2 and 0.8\,$\mu$m has been probed for the first time.\footnote{
Stecker \& Scully \cite{stecker:2009a} argued, however, that the derived limit has a low significance.}

\subsection{Discussion of the limits}

\begin{figure}
\includegraphics[viewport=20 45 485 520,clip,width=0.4\textwidth]{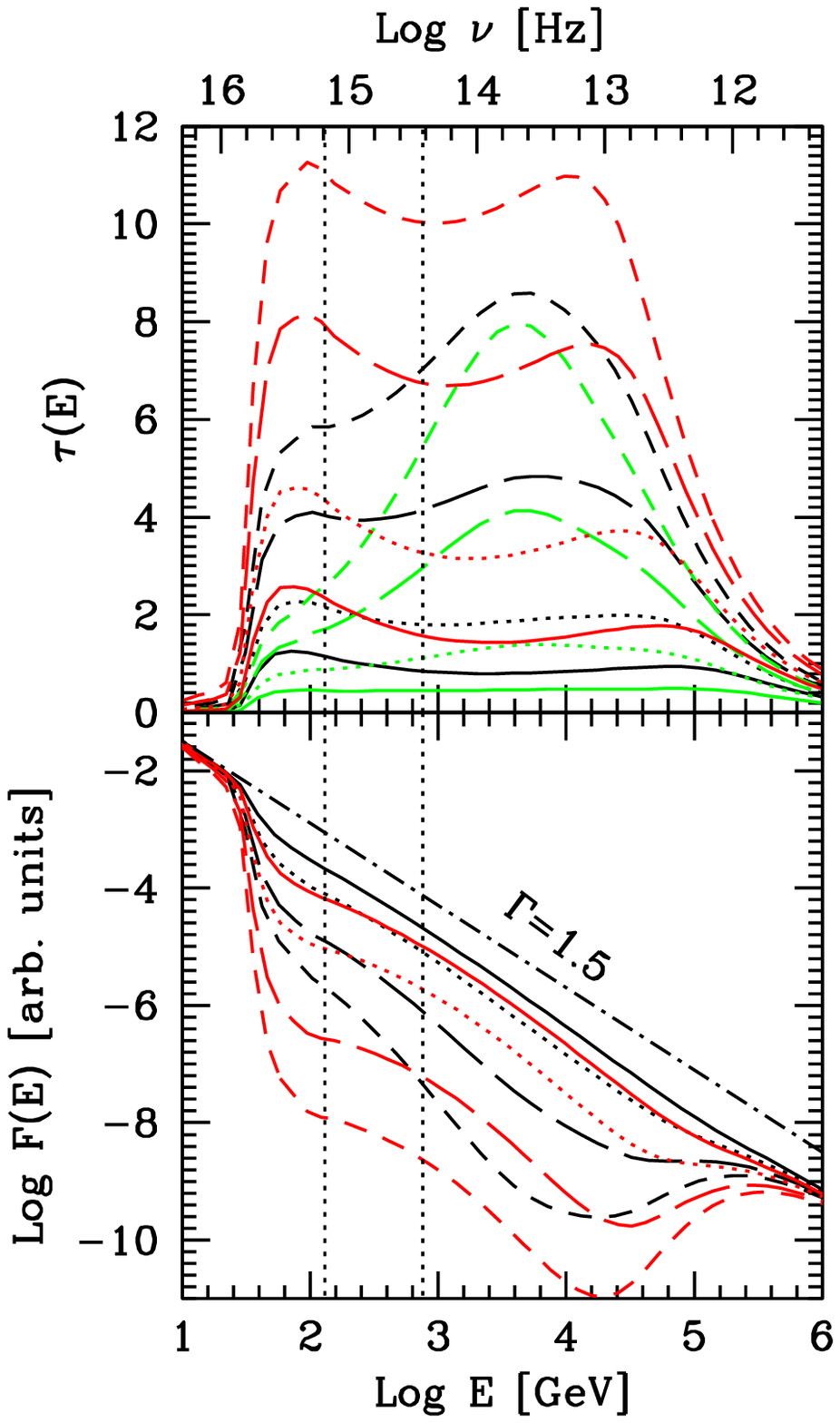}
\hfill
\includegraphics[width=0.5\textwidth]{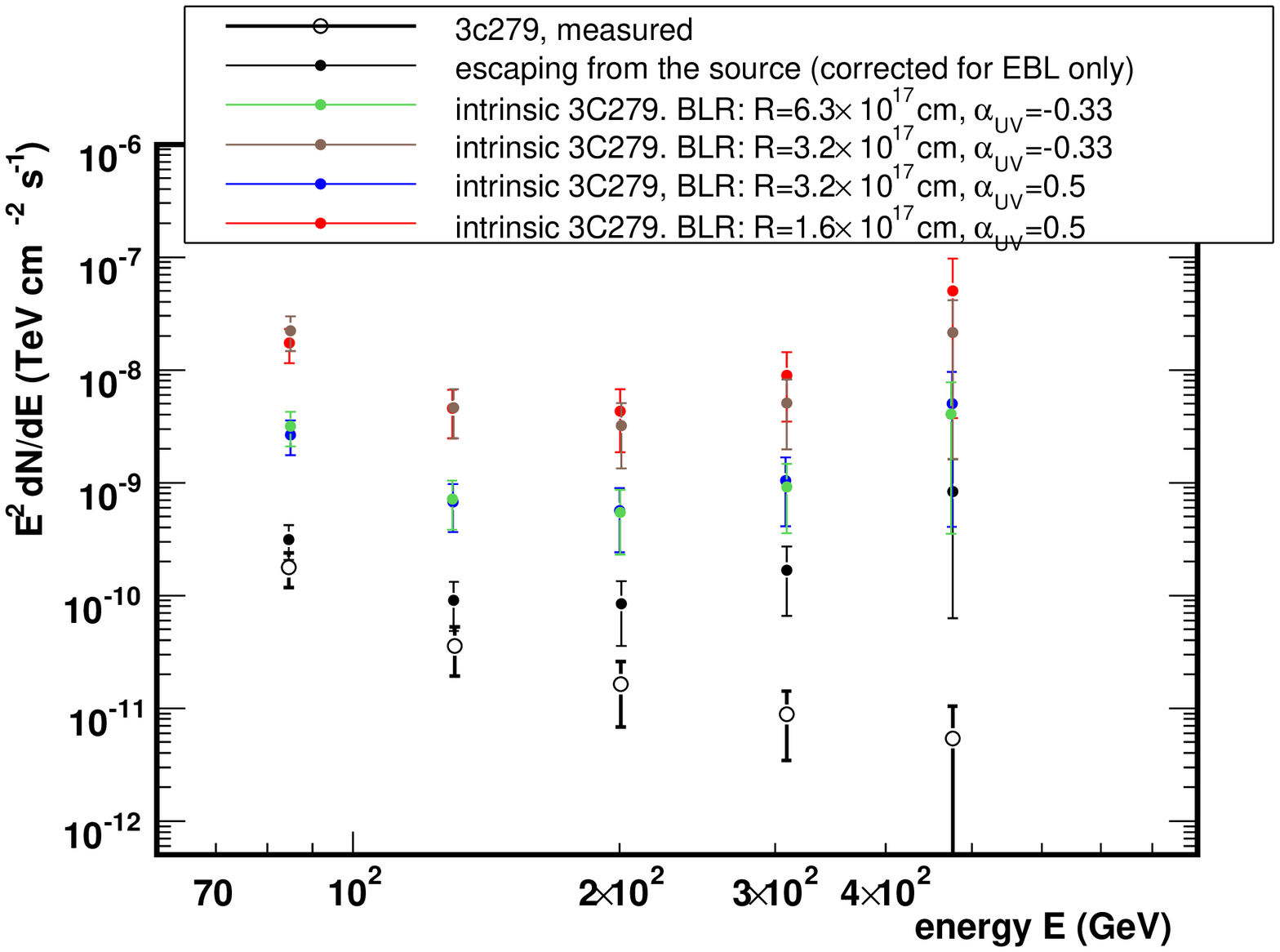}
\caption
{\label{fig:IntAbs} Internal absorption in case of 3C\,279.
{\it Upper left panel:} optical depth for absorption of
$\gamma$-rays for different parameters (see \cite{tavecchio:mazin:2008a} for details). 
{\it Lower left panel:} modification of
the intrinsic photon spectrum (assumed to be a power-law with slope
$\Gamma = 1.5$, black dashed-dotted line) by absorption. 
{\it Right panel:} 
Energy spectrum of 3C279. The measured spectrum (open
circles), EBL-corrected (filled black circles) and four different
scenarios for the internal absorption (in addition to the external
absorption) are shown. The EBL limits are identical or stronger than the ones 
in \cite{albert:2008:3C279:science} for all tested cases. Plots from
 \cite{tavecchio:mazin:2008a}.
} 
\end{figure}

Commenting the MAGIC result \cite{albert:2008:3C279:science} some doubts 
arose if the same criterion of  $\Gamma_{\rm int} = 1.5$ can be applied to 3C\,279
(e.g.\ Costamante, priv.\ communication, \citep{sitarek:2008a}). 
A possible problem is that  3C\,279
is not a blazar since (in a low flux state) it has prominent optical lines, identifying
low energy (optical - infrared) radiation fields, usually called ``broad line regions'', BLR. 
Indeed, the presence of BLR is required in most leptonic models describing the 
broad band emission (from radio through VHE \g-rays)
of 3C\,279. The BLR in front or within the emission region of VHE \g-rays  
leads to an internal absorption of these \g-rays, which modifies the intrinsic spectrum.
For some specific combinations of the geometry between the emission regions and the BLR
combined with a narrow band spectrum of the BLR,
local hardening of the intrinsic VHE \g-ray spectrum can be expected (e.g.\ \cite{aharonian:2008a}).
These question were addressed by \cite{tavecchio:mazin:2008a} in detail. The authors find
that, assuming a detailed BLR emission spectrum, 
for all plausible geometry combinations
no significant hardening of the intrinsic spectrum of 3C\,279 can  be expected
within the energy range observed by MAGIC (see Fig.~\ref{fig:IntAbs}). 
The authors also examined the EBL limits
including the effect of the internal absorption finding that identical or even harder EBL 
limits can be derived as compared to the ones obtained in \cite{albert:2008:3C279:science}. 

\begin{figure}
\includegraphics[width=0.5\textwidth]{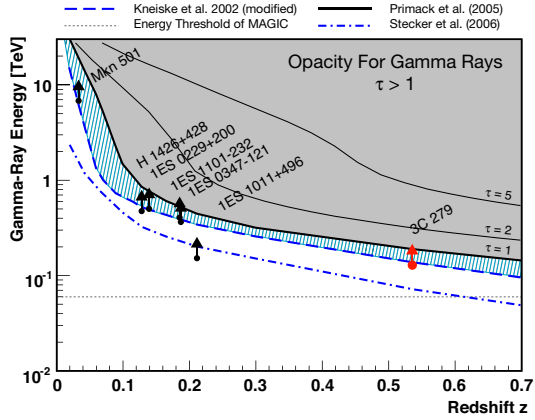}
\caption
{\label{fig:Horizon} Gamma-ray horizon.
IACTs probed the EBL density up to z = 0.536. The derived limits leave 
only a narrow region (indicated by a blue band) for EBL models. Figure from
\cite{albert:2008:3C279:science}.}
\end{figure}

Summarizing the status of the EBL constraints obtained by the IACTs, 
the following can be stated:
\begin{itemize}
 \item robust EBL upper limits are derived by different groups extending 
       from ultraviolet through mid infrared regimes; 
 \item the limits are close (at most factor of 2 higher) to the EBL low level inferred 
       from the resolved galaxies by $HST$, $ISOCAM$ and $Spitzer$; 
 \item this implies that instruments like $HST$, $ISOCAM$ and $Spitzer$ resolved most
       of the EBL sources;
 \item the resulting \g-ray horizon can be determined to lie within a narrow band
       between the upper limits from the IACTs and the low limits from the galaxy counts
       (see blue shaded band in Fig.~\ref{fig:Horizon});
 \item the limits disfavor several EBL models which imply a late peak in the star 
       formation history;
 \item the limits rule out a cosmological origin of the near infrared excess. 
\end{itemize}

However, the utilized exclusion criterion of the blazar intrinsic spectral index 
(being  $\Gamma_{\rm int} = 1.5$ or even $2/3$) can still be wrong. 
Even for an EBL model tuned to the level of the resolved galaxies 
\cite{franceschini:2008a}, the intrinsic spectra of several TeV blazars show the maximum
realistic hardness of 1.5. 
This can be related to the selection effect: only blazars with extremely
hard spectra can be detected because the flux of blazars with softer spectra falls below the
current sensitivity limit of the IACTs. 
Harder than expected intrinsic spectra of VHE \g-ray sources would imply either an
unnatural fine-tuning of low energy radiation fields inside the sources, 
different acceleration mechanisms of charged particles responsible for VHE \g-ray emission
or even new physics 
(e.g.\ violation of Lorentz invariance \cite{protheroe:2000a} or 
new particles \cite{deangelis:2007a}).  
Future observations with the $Fermi$ Gamma-ray observatory 
and new generation of IACTs such as H.E.S.S.\,II and MAGIC\,II will clarify the issue of 
hard intrinsic spectra due to a higher sensitivity of the instruments.

\section{Future perspective: Fermi, MAGIC\,II and H.E.S.S.\,II}

Currently, 26 extragalactic VHE \g-ray emitters have been detected with redshifts up to z = 0.536. 
Although some of the measured spectra have very good statistics, there is a general difficulty 
to distinguish between source intrinsic effects (e.g.\ cut-off corresponding to the highest energies 
of relativistic particles in the jet or internal absorption due to local low energy photon fields) 
and the attenuation imprint caused by the EBL.
The discovery of extragalactic sources with hard energy spectra by the HEGRA, H.E.S.S.\ and MAGIC
instruments allowed to set stringent limits on the EBL density in the near to mid infrared (NIR and MIR) 
regimes \cite{aharonian:2006:hess:ebl:nature,mazin:2007a}. 
The limits are, however, dependent on the 
underlying sources physics assumptions, which are still a matter of debate. We anticipate that 
MAGIC\,II and H.E.S.S.\,II will at least double the statistics of the sources but this will most likely not 
resolve the ambiguity between intrinsic effects and the EBL.
Probably, only CTA\footnote{planned next generation IACT array with a milli 
Crab sensitivity from 30 GeV through several 100 TeV} will be able to provide a sufficiently large sample (100++) 
of VHE \g-ray sources as well as high quality spectra for individual objects. 
For many of these sources, the spectral energy distribution (SED) will be determined at GeV energies 
(combining Fermi and lowest energies of the CTA measurements), which are much less affected by the absorption 
and, therefore, allowing for a study of intrinsic properties of the objects. We, therefore, 
anticipate that with CTA it will be possible to make robust predictions about the intrinsic spectrum above 
100 GeV, either for individual sources or particular source classes. 
The measured spectra will therefore be a unique probe of the EBL.
In the next 5 to 10 years following two main cosmological questions can be targeted: 
1) precision measurements of today's (i.e. at z = 0) EBL, especially in the ultraviolet to optical 
as well as in the mid to far infrared regimes, 
and 2) determine galaxy and stellar evolution properties by measuring the evolution of the EBL over redshift.


\begin{theacknowledgments}
  D.M.\ thanks the MAGIC collaboration and in particular MAGIC IFAE group in Barcelona for 
  the excellent support. D.M.\ also thanks Martin Raue for many fruitful discussions and Eckart Lorenz for a careful reading of the manuscript
  and useful comments. 
  This research was supported by a Marie Curie Intra European Fellowship within 
  the 7th European Community Framework Programme.
\end{theacknowledgments}



\bibliographystyle{aipproc}   

\bibliography{ebl_mazin.bbl}


\end{document}